\documentclass[aps, prl, amsfonts, amssymb, amsmath, floatfix, superscriptaddress, reprint, noeprint, longbibliography]{revtex4-2} 
\usepackage{times}
\usepackage{epsfig}
\usepackage{graphicx}
\usepackage{epstopdf}
\usepackage{bm}
\usepackage{color}
\usepackage[T1]{fontenc}
\usepackage[english]{babel}
\usepackage{microtype}

\PassOptionsToPackage{hyphens}{url}
\usepackage[breaklinks=true,colorlinks=true,citecolor=blue,urlcolor=blue,linkcolor=blue,pdfencoding=auto, psdextra]{hyperref}

\renewcommand{\vec}[1]{\mathbf{#1}}

\begin{document}

\title{Radiative Particle-in-Cell Simulations of Turbulent Comptonization in\\ Magnetized Black-Hole Coronae}

\author{Daniel~Gro\v selj}
\affiliation{Centre for mathematical Plasma Astrophysics, Department of Mathematics, 
KU Leuven, B-3001 Leuven, Belgium}
\affiliation{Department of Astronomy and Columbia Astrophysics Laboratory, 
Columbia University, New York, NY 10027, USA}
\author{Hayk~Hakobyan}
\affiliation{Computational Sciences Department, Princeton Plasma 
Physics Laboratory, Princeton, NJ 08540, USA}
\affiliation{Department of Physics and Columbia Astrophysics Laboratory, 
Columbia University, New York, NY 10027, USA}
\author{Andrei~M.~Beloborodov}
\affiliation{Department of Physics and Columbia Astrophysics Laboratory, 
Columbia University, New York, NY 10027, USA}
\affiliation{Max Planck Institute for Astrophysics, D-85741 Garching, Germany}
\author{Lorenzo~Sironi}
\affiliation{Department of Astronomy and Columbia Astrophysics Laboratory, 
Columbia University, New York, NY 10027, USA}
\author{Alexander~Philippov}
\affiliation{Department of Physics, University of Maryland, College Park, MD 20742, USA}

\date{\today}

\begin{abstract}

We report results from the first radiative particle-in-cell simulations of strong Alfv\' enic turbulence in 
plasmas of moderate optical depth. The simulations are performed in a local 3D periodic box and
self-consistently follow  the evolution 
of radiation as it interacts with a turbulent electron-positron
plasma via Compton scattering. 
We focus on the conditions expected in magnetized coronae of accreting black holes and
obtain an emission spectrum consistent with the observed hard state of Cyg X-1.
Most of the turbulence power is transferred directly to the photons via 
bulk Comptonization, shaping the peak of the emission around 100 keV. 
The rest is released into nonthermal particles, which generate the MeV spectral tail.
The method presented here shows promising potential for \emph{ab initio} modeling 
of various astrophysical sources and opens a window into a new regime of kinetic plasma turbulence.
\end{abstract}

\maketitle

{\em Introduction.---}Luminous accreting 
black holes at the cores of active galaxies and 
in X-ray binaries are some of the most prominent examples of high-energy electromagnetic emission \cite{Remillard2006,Padovani2017}. 
A particularly well-studied source is the binary Cyg X-1 \cite{Webster1972}, one of the 
brightest persistent sources of hard X-rays in the sky. The emission spectra of X-ray binaries 
are routinely observed in the soft and hard states \cite{Zdziarski2004},
with peak energies near 1 and 100 keV, respectively. The hard state is believed to originate 
from a hot ``corona'' of moderate optical depth \cite{Yuan2004,Fabian2015}, where 
the electrons Comptonize soft seed photons to produce the observed emission. 
The coronal electrons lose energy through inverse-Compton scattering, 
and therefore an energization process
is needed in order to balance the electron cooling. The nature of 
this process is unknown \cite{Poutanen2014}. 
In a number of proposed scenarios the electrons draw energy from magnetic fields.
The released magnetic energy is then channeled into 
bulk flows, nonthermal particles, and heat \cite{Socrates2004, Schnittman2013, Jiang2019, Liska2022, Kadowaki2015, Singh2015,
Khiali2015,Kaufman2016, Beloborodov2017,
Sironi2020,Sridhar2021,Mehlhaff2021,Sridhar2023}.

A fraction of the electron kinetic energy in black-hole coronae is likely 
contained in nonthermal particles \cite{Ghisellini1993,Zdziarski1993,Poutanen2014,Fabian2017},
which calls for a kinetic plasma treatment of their energization.
Among the various pathways leading to particle energization, not
only in black-hole accretion flows but in relativistic plasmas in general, turbulence has emerged as a prime candidate 
because it develops rather generically whenever the driving scale of the flow is much greater than the plasma
microscales \cite{Petrosian2012,Uzdensky2018}. Recent kinetic simulations explored relativistic turbulence 
in moderately \cite{Zhdankin2017, Zhdankin2018, Zhdankin2019,Wong2020} 
and strongly magnetized \cite{Comisso2018, Comisso2019,Nattila2022,Vega2022,Vega2022b,Bresci2022} nonradiative 
plasmas, and turbulent plasmas with a radiation reaction 
force on particles representing synchrotron or inverse-Compton 
cooling of optically thin sources \cite{Zhdankin2020, Comisso2021, Zhdankin2021, Nattila2021}. 
However, existing simulations do not apply to 
turbulence in black-hole coronae, which have
moderate optical depths.

In this Letter, we perform the first radiative kinetic simulations of 
turbulence in plasmas of \emph{moderate optical depth} 
and demonstrate that our method can
directly predict the observed emission from a high-energy astrophysical source.
As an example, we investigate here
the hard state of the archetypal source Cyg X-1. In 
the future, similar methods could be applied to study a variety of high-energy 
astrophysical systems.

{\em Method.---}We perform 3D simulations of driven
turbulence using the particle-in-cell 
(PIC) code \textsc{Tristan-MP v2} \cite{tristanv2}. 
All simulations 
employ for simplicity 
an electron-positron pair composition. 
The PIC algorithm is coupled with radiative transfer accounting for the injection of seed photons, photon escape, and
Compton scattering. The latter is resolved on a spatial grid composed of ``collision cells''
and incorporates Klein-Nishina cross sections \cite{Blumenthal1970, Rybicki1979}. 
The computational electrons (or positrons) and photons in a given collision cell are scattered 
using a Monte Carlo approach similar to \cite{Haugbolle2013,DelGaudio2020}, 
apart from a few technical adjustments 
described in Supplemental Material \footnote{See Supplemental Material, which 
includes Refs.~\cite{Stern1995, Politano1995, Dmitruk2004, Mininni2006, Sentoku2008, SantosLima2010, Eyink2011, Kowal2012b, Zhdankin2013, Vranic2015, Kadowaki2018, 
RodriguezRamirez2019, Kawazura2020, Lazarian2020, Sobacchi2021, Hinkle2021, Zhdankin2021b, Ripperda2022, Zhang2023}, 
for additional numerical details, simulation results, and discussions.}.
While the Compton scattering is modeled from first principles, we adopt for simplicity a more heuristic approach for photon injection and 
escape, as discussed below.

The simulation domain is a periodic cube of size $L$.
A mean magnetic field
$\vec B_0$ is imposed in the $z$ direction. 
We achieve a turbulent state
by continuously driving an external current in the form of a ``Langevin antenna'' \cite{TenBarge2014} 
that excites strong Alfv\' enic perturbations on the box scale \cite{Zhdankin2017,Groselj2019}.
The box is initially filled with
photons and charged particles in thermal equilibrium at temperature 
$T_0$. Given the lack of physical boundaries in the periodic box,
we implement a spatial photon escape 
by keeping track of how each photon diffuses from its initial injection location.
A given photon is removed from the box when it diffuses 
over a distance $l_{\rm esc}=L/2$
in any of the three Cartesian directions, so as to mimic escape from an
open cube of linear size $L$.
Each escaping photon is immediately replaced with a new seed photon, inserted at the location 
of the old particle, so that the total number of photons in the box remains constant.
The momenta of injected seed photons are sampled from an isotropic Planck spectrum
at the fixed temperature $T_0$.

Our setup has three key parameters: 
the pair plasma magnetization $\sigma_{\rm e} \equiv B_0^2 / 4\pi n_{\rm e0}m_{\rm e}c^2$, 
the ratio $n_{\rm ph0} / n_{\rm e0}$, and the Thomson optical depth $\tau_{\rm T} \equiv \sigma_{\rm T} n_{\rm e0}l_{\rm esc}$, 
where $n_{\rm e0}$ is the mean density of electrons and positrons,  $n_{\rm ph0}$ is 
the mean (X-ray and gamma-ray) photon density, and
$\sigma_{\rm T}$ is the Thomson cross section. 
Our fiducial simulation has $\sigma_{\rm e} = 2.5$, $n_{\rm ph0}/n_{\rm e0} = 250$, and
$\tau_{\rm T} = 1.7$. The choice of $\sigma_{\rm e}$ and  $\tau_{\rm T}$ mimics the conditions expected 
in black-hole coronae, which are believed to be 
strongly magnetized ($\sigma_{\rm e}\gtrsim 1$) and optically moderately thick ($\tau_{\rm T}\sim 1$) \cite{Merloni2001,Fabian2015}, whereas 
$n_{\rm ph0} / n_{\rm e0}$ is chosen such as to achieve an 
amplification factor $A\sim 10$ (defined below; see Eq.~\eqref{eq:A}), consistent with 
observations of hard states in X-ray binaries \cite{Beloborodov1999, Poutanen2014}.

Other parameters are chosen as follows.
The temperature of the seed photons is $T_0/m_{\rm e}c^2 = 10^{-3}$. 
We set the frequency and decorrelation rate 
of the Langevin antenna \cite{TenBarge2014} to $\omega_0 = 0.9 (2\pi v_{\rm A}/L)$
and $\gamma_0 = 0.5\omega_0$, respectively,
where $v_{\rm A}$ is the Alfv\' en speed. We define $v_{\rm A}= c [\sigma_{\rm e} / (1 + \sigma_{\rm e})]^{1/2}$.
The chosen strength of the antenna current results in a 
typical amplitude $\delta B \sim B_0$ for the large-scale fluctuating magnetic field. 
The simulation domain is resolved with $1280^3$ cells for the PIC scheme and $128^3$ 
collision cells for the Compton scattering. The size of the box  
is $L/d_{\rm e0} = 640$, where $d_{\rm e0} =  (m_{\rm e}c^2/4\pi n_{\rm e0}e^2)^{1/2}$ is the
 pair plasma skin depth. 
The time step for the PIC scheme and for Compton scattering 
is $\Delta t = 0.45 \Delta x/c$, where $\Delta x$ is the cell size of the PIC grid.
The plasma and radiation are each represented 
on average with eight macroparticles per cell of the PIC grid.
Additional simulations, numerical details, and discussions 
are included in Supplemental Material \cite{Note1}.

{\em Energy budget.---}Let us consider the energetics of the turbulent cascade. 
In steady state, the energy carried away by escaping radiation is balanced by the turbulence 
cascade power (cf. \cite{Uzdensky2018,Beloborodov2021}):
$n_{\rm ph0} (\overline E_{\rm esc} - \overline E_0) / t_{\rm esc} \simeq \delta B^2 / 4\pi t_0$,
where $\overline E_{\rm esc}$ and $\overline E_0$ are the mean energies of escaping and injected photons, respectively,
$t_{\rm esc} = ( \tau_{\rm T} + 1)l_{\rm esc}/c$ is the photon escape time associated with
diffusion over scale $l_{\rm esc}$, and 
$t_0 = l_0 / \delta v$ is the eddy turnover time at the
turbulence integral scale $l_0$ with velocity fluctuation
$\delta v$. Using $\delta v \approx (\delta B/B_0) v_{\rm A}$, we then obtain
\begin{align}
    \!\!\!\!A \simeq 1\!+\!\sigma_{\rm e} (\tau_{\rm T}\!+\!1)\! 
    \left(\!\frac{\overline E_0}{m_{\rm e}c^2}\!\right)^{\!\!\!-1}\!\!\!
    \left(\!\frac{n_{\rm e0}}{n_{\rm ph0}}\!\right)\!\left(\!\frac{v_{\rm A}}{c}\!\right)\!
    \left(\!\frac{\delta B}{B_0}\!\right)^{\!\!3}\!\left(\!\frac{l_{\rm esc}}{l _0}\!\right)\!,
    \label{eq:A}
\end{align}
where $A\equiv \overline E_{\rm esc} /\, \overline E_0$ is the amplification factor.
An effective electron temperature $\Theta_{\rm eff}$ can be obtained by balancing the
radiative cooling rate $\dot U_{\rm IC}$ with the power carried away by the escaping photons. 
To estimate $\dot U_{\rm IC}$ we assume for simplicity 
that the radiation field is isotropic, which is well satisfied when $\tau_{\rm T}\ll 1$; for $\tau_{\rm T}\sim 1$ moderate 
anisotropies may arise due to the scattering of photons by the large-scale bulk motions \cite{Zrake2019}.
In the regime of unsaturated Comptonization, relevant to black-hole coronae \cite{Shapiro1976}, we then have
$\dot U_{\rm IC} \simeq 4f_{\rm KN}\tau_{\rm T} n_{\rm ph0}\overline E_{\rm ph} \Theta_{\rm eff} c/ l_{\rm esc}$
(cf.~\cite{Beloborodov2021}), 
where $\Theta_{\rm eff} \equiv \overline{u^2}/3$, $u=\gamma\beta$ is the particle four-velocity 
in units of $c$, $f_{\rm KN}$ is a Klein-Nishina correction factor \cite{Moderski2005},
and $\overline E_{\rm ph}$ is the mean energy of a photon within the turbulent domain. 
Balancing $\dot U_{\rm IC}$ with $n_{\rm ph0}\overline E_{\rm esc} / t_{\rm esc}$ gives
\begin{align}
\Theta_{\rm eff}\simeq \frac{\overline E_{\rm esc}}{4\overline E_{\rm ph}f_{\rm KN}\tau_{\rm T}(\tau_{\rm T} + 1)}.
    \label{eq:Theta}
\end{align}
$\Theta_{\rm eff}$ is not to be confused with the proper plasma temperature. Rather, it should be regarded as 
a measure for the particle mean square four-velocity, which can 
include contributions from thermal, nonthermal, or bulk motions. 
For the time scale $t_{\rm IC} \simeq n_{\rm e0}\overline E_{\rm e}/ (n_{\rm ph0}\overline E_{\rm esc}/t_{\rm esc})$, on which 
the electron kinetic energy is passed to the radiation, we find
$t_{\rm IC} / t_0 \simeq (\overline E_{\rm e}/m_{\rm e}c^2) [\sigma_{\rm e}(\delta B/B_0)^{2}]^{-1}$, 
where $\overline E_{\rm e}$ is the mean kinetic energy per electron.
Finally, the radiative compactness \cite{Guilbert1983,Fabian2015} can be expressed 
as $\ell \simeq 4\tau_{\rm T}\sigma_{\rm e}(\delta B/B_0)^{3} (v_{\rm A}/c) (l_{\rm esc}/l_0)$ \cite{Note1}.

\begin{figure}[t]
\centering
\includegraphics[width=\columnwidth]{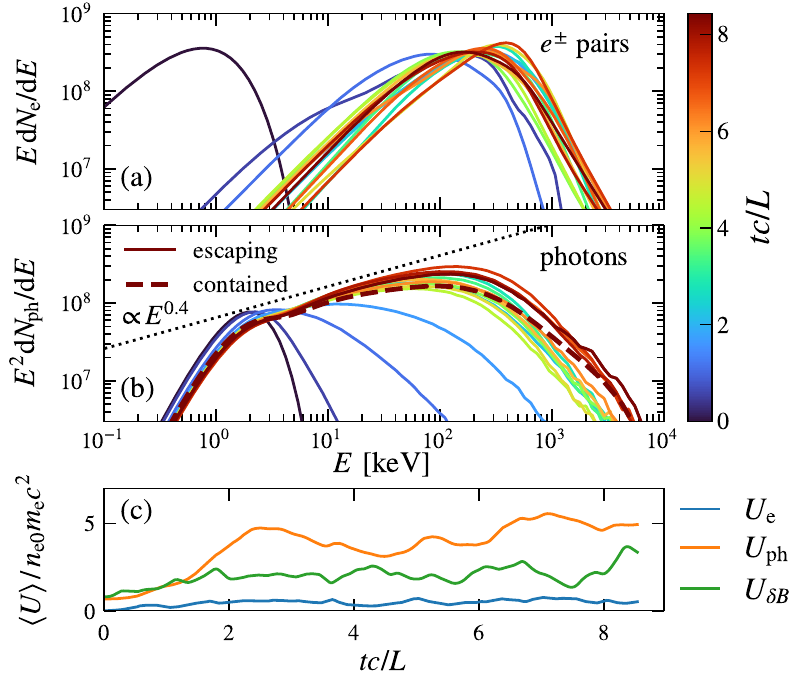}
\caption{Time evolution of the electron-positron (a) 
and escaping photon (b) energy spectrum, and the evolution of the box-averaged
plasma, radiation, and magnetic energy density (c). Different colors in panels (a) and (b)
represent the simulation time. 
Also shown is the spectrum of photons contained in the domain at the end of the
simulation (dashed red curve in panel (b)).\label{fig:1}}
\end{figure}

Fig.~\ref{fig:1} demonstrates the approach to a statistically steady turbulent state 
in our fiducial PIC simulation.
The charged particles and photons are energized by the turbulent cascade, reaching a quasi-steady state in 
roughly three light-crossing times $L/c$ \footnote{We ran simulations in smaller boxes 
up to $t c /L \approx 15$ and saw no signs of secular energy growth or decay beyond 
$t c /L \approx 3$. This supports our notion of a \emph{statistically} steady 
state in the larger but shorter fiducial run.}. Unless stated otherwise, the various 
statistical averages reported below represent the mean values over the quasi-steady state starting at
$t c/L \approx 3$ and extending until the end of the simulation.
The fully developed turbulent state exhibits random ``flaring'' activity associated with the buildup and
release of magnetic energy (Fig.~\hyperref[fig:1]{\ref*{fig:1}(c)}). The system is 
radiation-dominated and strongly magnetized
in the sense that both the box-averaged photon energy 
density $\langle U_{\rm ph}\rangle = n_{\rm ph0}\overline E_{\rm ph}$ 
and the fluctuating magnetic energy 
density $\langle U_{\delta B}\rangle = \langle\delta B^2\rangle/8\pi$ exceed the average kinetic
energy density $\langle U_{\rm e}\rangle = n_{\rm e0}\overline E_{\rm e}$ of pairs.

Consistent with observations \cite{Poutanen2014},
the escaping radiation spectrum exhibits in the statistically steady state
a photon index close to $\Gamma \approx 1.6$
between the photon injection energy of roughly 1 keV and the peak near 100 keV
(corresponding to $E^2{\rm d}N_{\rm ph}/{\rm d} E \propto E^{-\Gamma + 2} \sim E^{0.4}$ 
in Fig.~\hyperref[fig:1]{\ref*{fig:1}(b)}). For our simulation parameters
with $\overline E_0 / m_{\rm e}c^2 \approx 2.7\times 10^{-3}$, $\delta B/B_0 \approx 1$ 
and $l_{\rm esc}/l_0\approx 1.3$ \footnote{We define 
$l_0 = \pi (\int k_\perp^{-1}E_B(k_\perp){\rm d}k_\perp) / (\int E_B(k_\perp){\rm d}k_\perp)$, 
where $E_B(k_\perp)$ is the 1D
magnetic spectrum for wavenumbers perpendicular to $\vec B_0$.}, Eq.~\eqref{eq:A}
gives $A\approx 12$, in reasonable agreement with the typical 
value $A\approx 9$ measured in the simulation. 
The compactness $\ell \approx 19$ \cite{Note1}, which is 
comparable to the typical value $\ell\sim 50$ 
inferred for the hard state of Cyg X-1 \cite{Fabian2015}. The simulated
compactness is too low for a self-consistent
balance between pair creation and annihilation \citep{Note1}. Thus, 
a pair plasma composition is assumed here for computational convenience only. That our model does not include heavier 
ions is an aspect worth considering when comparing our results to magnetohydrodynamic (MHD) simulations.

The pairs develop over time a 
nonthermal spectrum (Fig.~\hyperref[fig:1]{\ref*{fig:1}(a)}) with mean kinetic energy per particle
$\overline E_{\rm e}/m_{\rm e}c^2 \approx 0.5$. The nonthermal tail ($E_{\rm e} \gtrsim 600$ keV) contains about 
30\% of the kinetic energy.
The effective temperature is raised by particles from the nonthermal tail to 
$\Theta_{\rm eff}\approx 0.6$. A thermal plasma with the same $\overline E_{\rm e}$ as measured 
in our simulation would have
a proper temperature $T_{\rm e}/m_{\rm e}c^2\approx 0.3$.
For reference, Eq.~\eqref{eq:Theta} predicts $\Theta_{\rm eff}\approx 0.2$ for 
our measured $\overline E_{\rm esc} /\,\overline E_{\rm ph}\approx 1.4$ 
and $f_{\rm KN} \approx 0.5$ \footnote{We use $f_{\rm KN} \approx {\langle }U_{\rm ph}{\rangle }^{-1}\!\int (1 + 4\overline{\gamma} \epsilon)^{-1.5}f_{\epsilon} {\rm d}\epsilon$, 
where $f_{\epsilon}$ is the photon spectral energy density, 
$\overline{\gamma} = \overline E_{\rm e}/m_{\rm e}c^2\!+\!1$, and $\epsilon = E_{\rm ph}/m_{\rm e}c^2$ \cite{Moderski2005}.}. 
For the cooling time scale we find $t_{\rm IC}/t_0 \approx 0.2$. Thus, the pairs pass their 
energy to the photons on a time scale shorter than the turbulent cascade time.

\begin{figure}[htb!]
\centering
\includegraphics[width=\columnwidth]{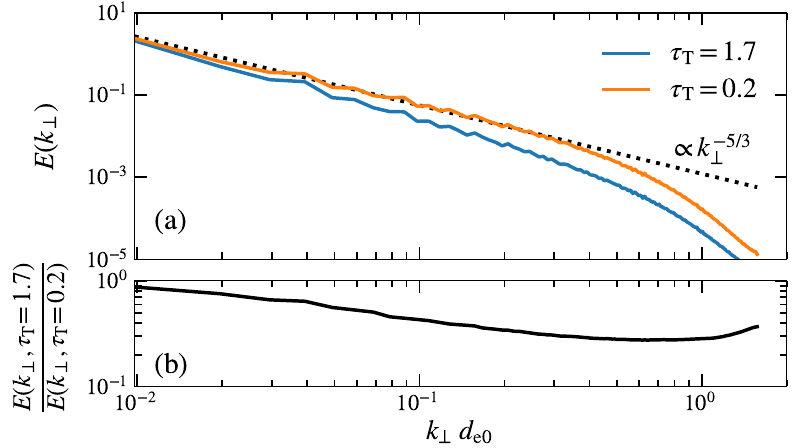}
\caption{\label{fig:2}1D power spectra $E(k_\perp)$ of the turbulence energy as a 
function of the wavenumber $k_\perp$ perpendicular to $\vec B_0$ for 
$\tau_{\rm T}=1.7$ and $\tau_{\rm T}=0.2$ (a). Panel (b) shows the
ratio of the turbulent spectra from the two simulations.}
\end{figure}

{\em Emission mechanism.---}The Comptonization of photons 
can occur through internal 
or bulk motions. In the fast cooling regime ($t_{\rm IC} < t_0$), 
a fraction $f_{\rm bulk}$ of the turbulence power is passed to the photons 
via bulk Comptonization \emph{before} the cascade reaches the plasma microscales, 
leading to radiative damping of the turbulent flow \cite{Thompson1994, Thompson2006, Zrake2019}. 
This is demonstrated in Fig.~\ref{fig:2}, which shows the turbulence 
energy spectra $E(k_\perp)$, defined as the sum of magnetic, electric, and bulk kinetic energy density 
spectra \footnote{The kinetic energy density spectrum is obtained as the power spectrum of 
$\vec w = \left[n_{\rm e}m_{\rm e}c^2\gamma^2_{\rm bulk}/(\gamma_{\rm bulk} + 1)\right]^{1/2}\boldsymbol{\beta}_{\rm bulk}$, such that 
$|\vec w|^2 = (\gamma_{\rm bulk} - 1)n_{\rm e} m_{\rm e}c^2$.}. 
The spectrum $E(k_\perp)$ from our run with $\tau_{\rm T}=1.7$ is compared against the result obtained from 
a simulation with $\tau_{\rm T}=0.2$ but otherwise identical parameters. The spectra
extend from the injection scale ($k_\perp d_{\rm e0}\sim 0.01$) into the kinetic range 
($k_\perp d_{\rm e0}\gtrsim 1$), where the cascaded energy converts into plasma internal 
motions. Over the MHD range ($k_\perp d_{\rm e0}\ll 1$) 
the turbulence spectrum for $\tau_{\rm T}=0.2$ exhibits
a slope consistent with a classical cascade where  
$E(k_\perp)\propto k_{\perp}^{-5/3}$ \cite{Goldreich1995,Thompson1998}, while for $\tau_{\rm T} = 1.7$ the radiative damping 
becomes strong enough to steepen the spectrum (Fig.~\hyperref[fig:2]{\ref*{fig:2}(a)}). 
This can be considered an example 
for how radiative effects render the turbulence spectra non-universal.

The steepening of the turbulence spectrum in our fiducial simulation with $\tau_{\rm T}=1.7$
is related to the power lost via bulk Comptonization as follows. In the MHD range, it may be assumed 
that $\Pi_{k_\perp}\sim {\mathcal F}_0 - {\mathcal D}^{\rm rad}_{k_{\perp}}$, where 
$\Pi_{k_\perp}$ is the turbulent energy flux to perpendicular wavenumbers larger than $k_\perp$,
${\mathcal F}_0$ is the external driving confined to the 
wavenumber $k_0\ll k_{\perp}$, and ${\mathcal D}^{\rm rad}_{k_{\perp}}$ 
is the radiative dissipation rate between $k_0$ and $k_{\perp}$. 
Since a fraction $f_{\rm bulk}$ of the cascade power is lost to radiation, we 
have ${\mathcal D}^{\rm rad}_{k_{\max}} \sim f_{\rm bulk} {\mathcal F}_0$, and 
so $\Pi_{k_\perp} / \Pi_0 \sim 1  - f_{\rm bulk} {\mathcal D}^{\rm rad}_{k_{\perp}} / {\mathcal D}^{\rm rad}_{k_{\max}} $, 
where $\Pi_0 \sim {\mathcal F}_0$ is the energy flux in the absence of damping. The flux can be 
approximated as $\Pi_{k_\perp} \propto k_\perp^{2 + \alpha} E(k_\perp)^{1 + \alpha}$, 
with $\alpha = 1/2$ for the Goldreich-Sridhar turbulence model \cite{Goldreich1995,Thompson1998}. 
There follows the estimate
\begin{align}
E(k_\perp)/E_0(k_\perp) \sim \left(1 - f_{\rm bulk} 
{\mathcal D}^{\rm rad}_{k_{\perp}}/{\mathcal D}^{\rm rad}_{k_{\max}}\right)^{\frac{1}{1+\alpha}},
\label{eq:damping}
\end{align}
where $E_0(k_\perp)$ is the spectrum in the absence of significant damping.
At the tail of the MHD range ($k_\perp d_{\rm e0}\sim 0.5$), we have
$f_{\rm bulk} \sim 1 - (E(k_\perp) / E_0(k_\perp))^{1 + \alpha}$, which can be taken as a
proxy for measuring $f_{\rm bulk}$. We substitute for $E_0(k_\perp)$ 
the spectrum obtained for $\tau_{\rm T}=0.2$ and estimate from  
Fig.~\hyperref[fig:2]{\ref*{fig:2}(b)} that 
$E(k_\perp) / E_0(k_\perp) \approx 0.3$ near $k_\perp d_{\rm e0}\approx 0.5$ \footnote{Bulk Comptonization is a meaningful concept as long as 
the turbulence is fluidlike, which is marginally satisfied
up to the transition into the kinetic range.
Thus, we measure $E(k_\perp)/E_0(k_\perp)$ just slightly below $k_\perp d_{\rm e0}\sim 1$.}, 
indicating that
roughly $f_{\rm bulk}\approx 80$\% (using $\alpha = 1/2$) of the total cascade power is 
passed to the photons via bulk Comptonization. The turbulent flow is dominated by 
motions transverse to the magnetic field \cite{Cho2003}, which renders the emission anisotropic.
The intensity of Comptonized radiation escaping parallel to $\vec B_0$ is about 3 times lower than 
the intensity emitted perpendicular to the mean magnetic field. Note that
efficient bulk Comptonization is generally expected when the particles cool quickly ($t_{\rm IC} < t_0$).
Guided by our simulation, we can give a simple estimate of $f_{\rm bulk}$ for $t_{\rm IC} < t_0$ as
$f_{\rm bulk} \sim 1 - t_{\rm IC}/t_0\sim 1 - (\overline E_{\rm e}/m_{\rm e}c^2) [\sigma_{\rm e}(\delta B/B_0)^{2}]^{-1}$, which
connects bulk Comptonization to the high-$\sigma_{\rm e}$ regime.

\begin{figure*}[htb!]
\centering
\includegraphics[width=\textwidth]{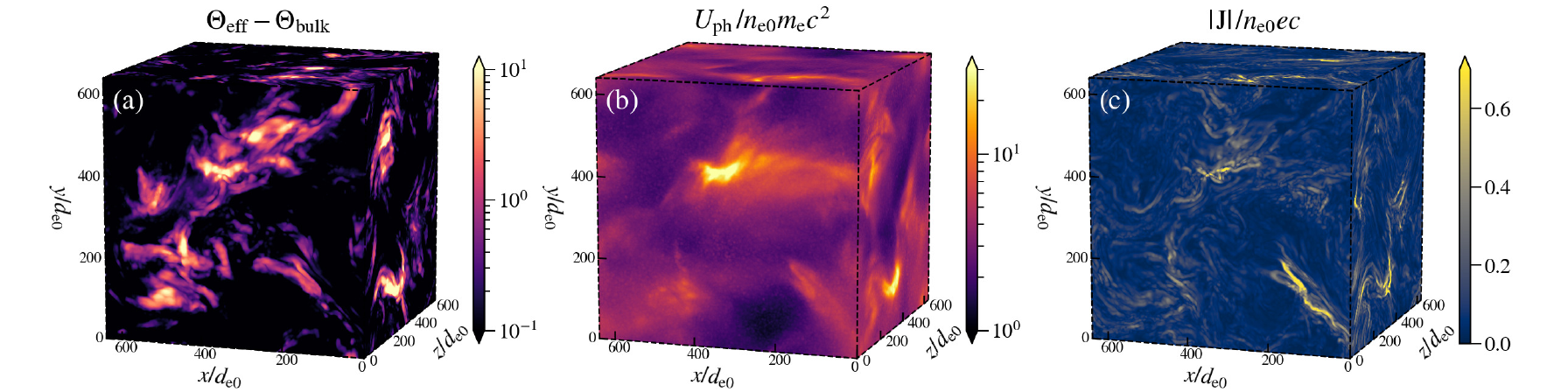}
\caption{Spatial structure of $\Theta_{\rm eff} - \Theta_{\rm bulk}$ (a), $U_{\rm ph}$ (b), and $\left|\vec J\right|$ (c),
where $\Theta_{\rm eff}$ is the effective plasma temperature, $\Theta_{\rm bulk}$ is the
``temperature'' of turbulent bulk motions, $U_{\rm ph}$ is 
the photon energy density, and $\left|\vec J\right|$ is the magnitude of the plasma electric current.\label{fig:visuals}}
\end{figure*}

Efficient bulk Comptonization implies that the plasma is essentially cold
and its effective temperature is 
close to the ``temperature'' of turbulent bulk 
motions $\Theta_{\rm bulk} \equiv u_{\rm bulk}^2/3$ \cite{Socrates2004},
where $u_{\rm bulk}^2 = \beta^2_{\rm bulk} / (1 - \beta^2_{\rm bulk})$ is the 
squared bulk four-velocity in units of $c^2$. In the simulation 
with $\tau_{\rm T}=1.7$ we find on average $\Theta_{\rm bulk} / \Theta_{\rm eff} \approx$ 50\%.
In Fig.~\ref{fig:visuals} we visualize the local difference $\Theta_{\rm eff} - \Theta_{\rm bulk}$ 
at time $t c/L = 6$ in our fiducial simulation. 
For reference, we also show the structure of the photon energy density
and the magnitude of the plasma electric current. Over much of the volume the plasma is indeed cold, 
in the sense that at most locations the difference $\Theta_{\rm eff} - \Theta_{\rm bulk}$ is very moderate. 
In a small fraction of the volume, typically near electric current sheets, the turbulent energy 
is intermittently released into internal motions,
giving rise to ``hot spots'' with $\Theta_{\rm eff} - \Theta_{\rm bulk} \gtrsim 1$. The hot spot formation requires a 
rapid form of energy release in order to outpace the fast cooling. One promising candidate 
is magnetic reconnection \cite{Note1}, which is known to promote particle energization 
magnetically dominated MHD \cite{Lazarian1999, DalPino2005, Kowal2009, Kowal2012, Lazarian2012, 
Eyink2013, DalPino2015, Takamoto2015, delValle2016, Beresnyak2016}
and kinetic \cite{Comisso2019,Comisso2021,Guo2021,Zhang2021,Chernoglazov2023} turbulent plasmas.

\begin{figure}[htb!]
\centering
\includegraphics[width=\columnwidth]{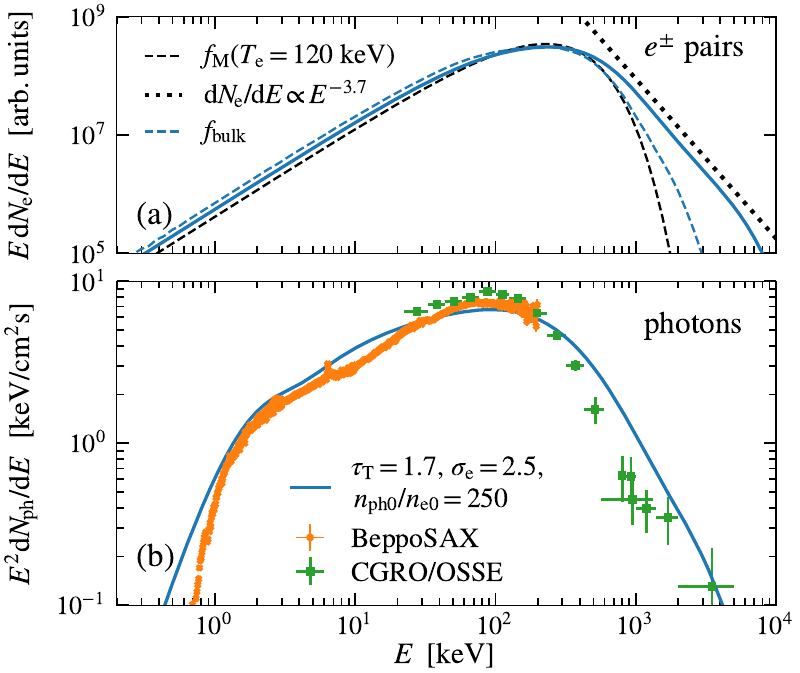}
\caption{\label{fig:4}Energy spectra of electron-positron pairs (a) and
of the escaping radiation (b), overplotted with observations of the hard state
in Cyg X-1 from BeppoSAX \cite{Salvo2001,Frontera2001} 
and CGRO/OSSE \cite{McConnel2002}. The emission spectra are normalized with
respect to OSSE. Dashed blue curve in panel (a) shows the energy distribution due to 
bulk motions alone. The dashed black curve shows a Maxwellian distribution fitted below 400 keV.}
\end{figure}

{\em Observational implications.---}Fig.~\ref{fig:4} shows the
spectra from our fiducial PIC simulation, time-averaged over steady state, 
together with observations of Cyg X-1 in the hard state. 
The obtained emission spectrum closely resembles the observations.
Differences between our simulation and observations
are seen below 1 keV, where the observed spectrum is attenuated by absorption, 
between 10 keV and the peak, and around 1 MeV. 
We do not include the additional radiation component
that is Compton-reflected from the disk \cite{Zdziarski2004}, which 
affects the spectrum in the range between roughly 10 keV and the peak. 
Regarding the MeV tail, we note that the inclusion of synchrotron 
cooling \cite{Poutanen2014} and pair creation \cite{Svensson1984,Svensson1987} 
could soften the tail. Simulations with electron-ion compositions, 
pair creation and annihilation, and/or synchrotron emission 
can further constrain the physical conditions required to reproduce the observed MeV tail.

The strongly magnetized regime ($\sigma_{\rm e} \gtrsim 1$)  explored here corresponds to a 
radiatively compact corona ($\ell \gtrsim 10$) located roughly within 10 gravitational radii from the
black hole \cite{Note1}.
A natural feature of our model 
is the formation of a nonthermal electron tail (Fig.~\hyperref[fig:4]{\ref*{fig:4}(a})), which shapes the MeV emission.
The distribution due to bulk motions alone
(dashed blue curve in Fig.~\hyperref[fig:4]{\ref*{fig:4}(a})) is significantly less nonthermal
than the full distribution (solid blue curve), which implies that the 
nonthermal tail is mostly contributed by internal motions. We ran an additional simulation at $\sigma_{\rm e}=0.1$ 
and found in comparison to our fiducial run a weaker nonthermal electron tail \citep{Note1}. 
The results may also depend on the type of turbulence driving (e.g., the low-amplitude regime with $\delta B\ll B_0$ 
is less favorable for the production of nonthermal particles \citep{Note1, Comisso2018, Nattila2022}).

{\em Conclusions.---}We performed the first PIC simulations of plasma turbulence that self-consistently follow the evolution of 
radiation via Compton scattering. 
Our simulations focus on the conditions expected in magnetized coronae of accreting black holes \cite{Zdziarski2004,Fabian2015},
which have moderate optical depths and experience fast radiative cooling. Similar conditions can also arise in jets of gamma-ray
bursts \cite{Thompson1994, Rees2005, Thompson2006, Ryde2011,Beloborodov2017c, Zrake2019}.
We obtain a spectrum of escaping X-rays similar to the observed hard-state spectrum of Cyg X-1, 
thus demonstrating that kinetic turbulence is a viable mechanism for the energization of electrons in black-hole coronae.

While the Compton scattering between the turbulent kinetic plasma and the radiation is treated self-consistently,
we note that our present setup is still subject to a number of limitations. We do not model the emission of soft photons, pair creation, annihilation, or
the global structure of the extended corona and the accretion disk. Instead, we adopt a local 3D periodic box approximation with a fixed average pair number density and with soft photon injection matching photon escape to sustain a fixed photon-to-electron ratio.
A complete understanding of the X-ray emission from black-hole coronae may require a global kinetic model with detailed radiative transfer, which is presently lacking. 
Existing global models based on MHD simulations (e.g., \citep{Schnittman2013,Jiang2019,Liska2022}) suggest that the properties of the observed X-rays depend not 
only on the mechanism of local energy release into radiation, but also on the geometric shape and multiphase structure of the corona.

In our local model, the emission is produced
via Comptonization in a plasma energized by large-amplitude ($\delta B\sim B_0$) Alfv\' enic turbulence.
For a strongly magnetized plasma,
we find that most of the turbulence power is directly passed to the photons through 
bulk Comptonization. The rest is channeled into nonthermal particles 
at localized hot spots. For computational convenience, our simulations employ a pair plasma composition.
The nature of turbulent
Comptonization in electron-ion plasmas could differ from that in pair plasmas \citep{Note1}. An 
important parameter is the fraction of turbulence power channeled into ion heating, which needs to be investigated with dedicated simulations.
We also show that turbulent Comptonization manifests itself through non-universal 
turbulence spectra.  As such, our simulations give a glimpse into a new 
regime of kinetic turbulence in radiative plasmas of moderate optical depth.

\begin{acknowledgments}

We acknowledge helpful discussions with L.~Comisso, J.~N\" attil\" a, V.~Zhdankin, B.~Ripperda, and R.~Mushotzky. We also 
thank N.~\mbox{Sridhar} for his assistance in obtaining observational data for Cyg X-1.
D.G.~is supported by the Research Foundation -- Flanders (FWO) Senior Postdoctoral 
Fellowship 12B1424N. D.G.~was also partially supported by the U.S.~DOE Fusion Energy 
Sciences Postdoctoral Research Program administered by ORISE for the DOE. ORISE is managed by ORAU under DOE contract 
DE-SC0014664. All opinions expressed in this paper are the authors’ and do not necessarily reflect the policies and views of
DOE, ORAU, or ORISE. L.S.~acknowledges support by the Cottrell Scholar Award. L.S.~and D.G.~were 
also supported by NASA ATP grant 80NSSC20K0565. A.M.B.~is supported by NSF grants AST-1816484 and AST-2009453, NASA grant \mbox{21-ATP21-0056}, and Simons Foundation grant 446228.
A.P.~was supported by NASA ATP grant 80NSSC22K1054. The work was supported by a grant from the Simons Foundation (MP-SCMPS-00001470, to L.S.~and A.P.).
An award of computer time was provided by the INCITE program. This research used resources of the Argonne 
Leadership Computing Facility, which is a DOE Office of Science User Facility supported under contract DE-AC02-06CH11357.
Simulations were additionally performed on NASA Pleiades (GID s2754). This research was facilitated by the Multimessenger Plasma Physics Center (MPPC), NSF grant PHY-2206607. 
\end{acknowledgments}


%


\pagebreak
\clearpage
\widetext
\begin{center}
{\large \textbf{Supplemental Material for ``Radiative Particle-in-Cell Simulations of Turbulent Comptonization in\\ Magnetized Black-Hole Coronae''}}
\end{center}
\setcounter{equation}{0}
\setcounter{figure}{0}
\setcounter{table}{0}
\setcounter{page}{1}
\setcounter{section}{0}

\section{Radiative compactness and electron-positron pair balance}
\label{sec:compactness}

The particle composition of black-hole coronae may be electron-ion or pair dominated, 
depending on the radiative compactness $\ell$ and on the energy distribution 
of the Comptonizing electrons \cite{Guilbert1983,Svensson1984,Svensson1987,Stern1995,Fabian2015,Fabian2017}. In particular, pair dominated states are 
associated with radiatively compact sources ($\ell\gg 1$). In a local cubic slab of linear size $L$ the radiative compactness may be defined 
as $\ell = P_{\rm diss} \sigma_{\rm T} / L m_{\rm e}c^3$ (e.g., \cite{Stern1995}), where $P_{\rm diss}$ is the dissipated power transferred to the radiation. 
For our simulation setup we estimate $P_{\rm diss} \simeq L^3 \delta B^2 / 4 \pi t_0 $. 
Using $L = 2l_{\rm esc}$  we obtain
\begin{align}
    \ell \simeq 4\tau_{\rm T}\sigma_{\rm e}\!\left(\delta B/B_0\right)^3\!\left(v_{\rm A}/c\right)\!\left(l_{\rm esc}/l_0\right).
    \label{eq:ell}
\end{align}
This shows that a high magnetization ($\sigma_{\rm e}\gtrsim 1$) translates into a large radiative compactness in our model, consistent with previous arguments in support 
of a magnetically dominated corona \cite{Merloni2001}. In our fiducial simulation we obtain $\ell \approx 19$, 
which is comparable to the  typical value $\ell\sim 50$ inferred for Cyg X-1 \cite{Fabian2015}. It is also 
consistent with the broader range $\ell \sim 1 - 100$ representative of black-hole coronae in active galactic nuclei \cite{Hinkle2021}. 
The regime $\ell \lesssim 1$ can be identified with $\sigma_{\rm e}\ll 1$. Here, the electrons are only mildly
nonthermal, their cooling is slow, and the emission mechanism is essentially thermal Compotonization (see Sec.~\ref{sec:low_sigma}). 
Using Eq.~(1) from the main Letter, we can 
alternatively express $\ell\sim 4 \bigl(n_{\rm ph0}/n_{\rm e0}\bigr)\bigl(\overline E_{\rm esc}/m_{\rm e}c^2\bigr)\bigl[\tau_{\rm T}/(\tau_{\rm T} + 1)\bigr]$,
which shows that $\ell$ scales in proportion to $n_{\rm ph0}/n_{\rm e0}$.
In global accretion models, one can also estimate 
$\ell \sim 4\pi (m_{\rm p}/m_{\rm e}) \tilde R^{-1}\tilde L$ \cite{Fabian2015},
where $m_{\rm p}$ is the proton mass, $\tilde L$ is the luminosity in units
of the Eddington luminosity, and $\tilde R$ is a characteristic size of the
source in units of the gravitational radius. A luminous black hole accreting
at a few per cent of the Eddington limit has roughly $\ell \gtrsim 10$ when 
$\tilde R \lesssim 10$. Thus, the high magnetization regime
($\sigma_{\rm e}\gtrsim 1$) of our local model corresponds to a radiatively compact corona
located roughly within 10 gravitational radii or so from the black hole.

It is worth estimating how close (or far) is our fiducial PIC simulation from a state of pair balance, 
where pair creation is balanced by annihilation. A simple but direct estimate can be given based on a set of 
analytic approximations \cite{Svensson1987,Beloborodov2017} for the time scales of pair creation, $t_{\gamma\gamma} = n_{\rm e^+} / \dot n_{\gamma\gamma}$, and 
annihilation, $t_{\rm ann} = n_{\rm e^+} / \dot n_{\rm ann}$. In our notation, these can be expressed as:
\begin{align}
    t_{\gamma\gamma} / t_{\rm esc} & \sim \frac{n_{\rm e0}^2 / n_{\rm ph0}^2}{2\eta f_1^2\,\tau_{\rm T}(1+\tau_{\rm T})}, &  
    t_{\rm ann} / t_{\rm esc} & \sim \frac{16}{3\tau_{\rm T}(1+\tau_{\rm T})}, &
    \label{eq:pp_and_ann}
\end{align}
where $f_1$ is the fraction of photons with energies $E_{\rm ph}\geq m_{\rm e}c^2$, $\eta\sim 0.1$ \cite{Svensson1987}, 
and $t_{\rm esc}$ is the photon escape time. This gives $t_{\gamma\gamma} / t_{\rm esc}\sim 30$ and
$t_{\rm ann} / t_{\rm esc} \sim 1$ for our measured $f_1 \approx 7.2\times 10^{-4}$. Taken at face value, the estimated rate of 
pair production is too slow to maintain an optical depth $\tau_{\rm T}\sim 1$. The electron-positron particle composition 
is chosen here mainly for simplicity, given that an equivalent simulation with heavier ions is computationally infeasible at present. 
On the other hand, it should be also noted that the time scale of pair creation in Eq.~\eqref{eq:pp_and_ann} 
is only a crude estimate, obtained for an isotropic and homogeneous radiation field. Moreover, $t_{\gamma\gamma}$ is very 
sensitive to the number of photons in the MeV tail, such that even moderate changes in the input parameters 
can lead to large differences in the pair creation rate. According to the crude estimate from above, 
a pair balanced state ($t_{\gamma\gamma} \sim t_{\rm ann}$) might be obtained in our model for 
a compactness of the order of 100 or so.
Radiative PIC simulations with self-consistent pair creation and annihilation are required 
to accurately constrain the parameters of pair-balanced states for the turbulent Comptonization model.

\section{Plasma-dominated regime with slow radiative cooling (\texorpdfstring{$\sigma_{\rm e}\ll 1$}{sigma})}
\label{sec:low_sigma}

\begin{figure*}[htb!]
\centering
\includegraphics[width=0.49\textwidth]{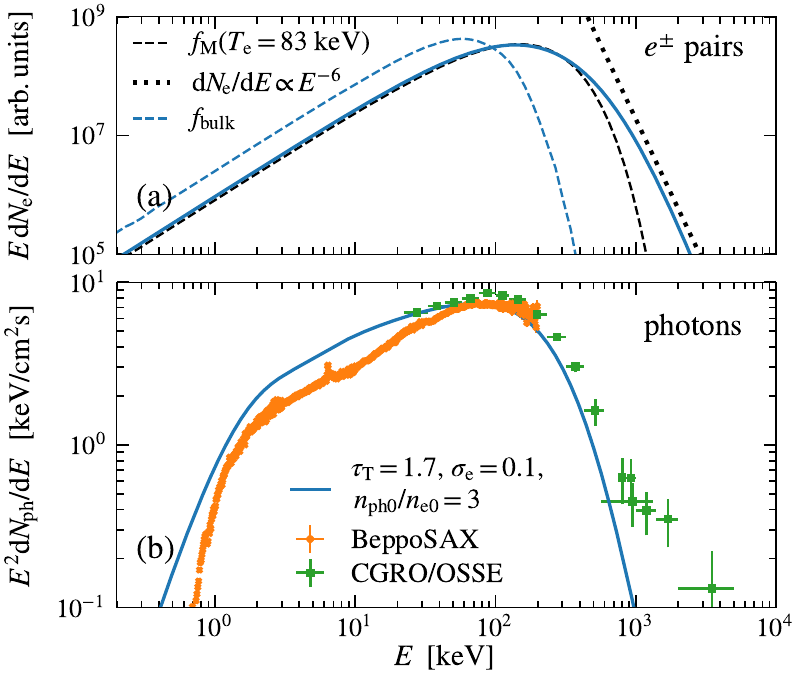}
\includegraphics[width=0.49\textwidth]{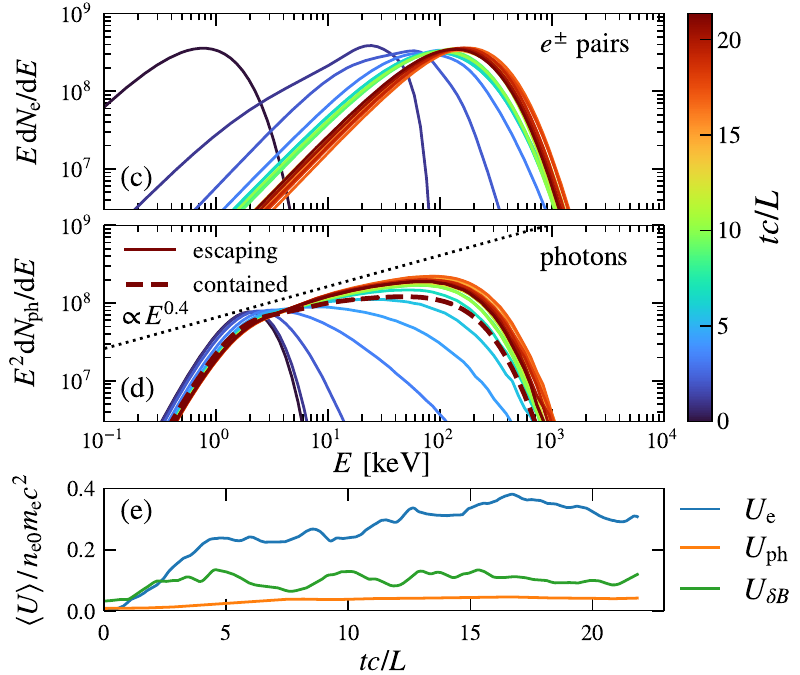}
\caption{\label{figsupp:lowsig} Results from a PIC simulation performed in the low-$\sigma_{\rm e}$
regime with slow radiative cooling.}
\end{figure*}

Our Letter focuses on the application of turbulent cascades to sources with fast radiative cooling of electrons, 
as appropriate for magnetized coronae of accreting black holes \cite{Merloni2001,Fabian2015}. This can be contrasted with the slow cooling regime, which
corresponds to $\sigma_{\rm e}\ll 1$ in our model. In Fig.~\ref{figsupp:lowsig} we show for reference results from 
a simulation with $\sigma_{\rm e} = 0.1$. In order to achieve an amplification factor $A\sim 10$ at the reduced
value of $\sigma_{\rm e}$ we set $n_{\rm ph0} / n_{\rm e0} = 3$ (according to
Eq.~(1) from the main Letter). Other simulation parameters match those reported in the main Letter.

As shown in Fig.~\hyperref[figsupp:lowsig]{\ref*{figsupp:lowsig}(e)}, the system evolves toward a state 
where the particle kinetic energy density ($U_{\rm e}$) dominates over the magnetic ($U_{\delta B}$) and
radiation ($U_{\rm ph}$) energy density. In the quasi-steady state (starting around $\approx 7 L/c$), we 
measure $A\approx 7$ and $\ell \approx 0.3$, which is inconsistent with the 
typical range $\ell \sim 1 - 100$ inferred from observations of accreting black holes. 
As expected, the turbulent cascade is
in the slow cooling regime with $t_{\rm IC} / t_0 \approx 3$. Moreover, the 
electron energy distribution (Fig.~\hyperref[figsupp:lowsig]{\ref*{figsupp:lowsig}(a)}) features only a 
mild nonthermal tail (at $E_{\rm e}\gtrsim$ 600 keV), 
which contains less than 10\% of the kinetic energy. As a result, the escaping photon spectrum shows
no significant emission in the MeV range (Fig.~\hyperref[figsupp:lowsig]{\ref*{figsupp:lowsig}(b)}).

\section{Turbulent Comptonization in electron-ion plasmas}

All simulations presented here employ for simplicity an electron-positron pair particle composition. 
However, the composition of the coronal plasma may be dominated by ions and electrons rather than pairs (see Sec.~\ref{sec:compactness} for a discussion).
If the particle 
composition is dominated by electrons and heavier ions (i.e., protons) an additional parameter enters the
problem: the fraction $q_{\rm i}$ of turbulence power channeled into ion heating. The parameter $q_{\rm i}$ 
needs to be determined from kinetic models and simulations; it has been extensively studied in 
nonradiative simulations (e.g., \cite{Zhdankin2019, Kawazura2020}) and in one set of simulations with inverse-Compton cooling of
relativistically hot electrons \cite{Zhdankin2021}. The ion heating fraction in the regime relevant to this work 
($\tau_{\rm T}\sim 1$, fast Compton cooling, and mildly relativistic electrons) has not been investigated.
Our scalings, obtained for a pair plasma, 
can be easily adapted for the electron-ion case. We find
\begin{align}
    A & \sim   1 + (1 - q_{\rm i})\sigma_{\rm e} (\tau_{\rm T} + 1) 
    \left(\!\frac{\overline E_0}{m_{\rm e}c^2}\!\right)^{\!\!\!-1}\!\!
    \left(\!\frac{n_{\rm e0}}{n_{\rm ph0}}\!\right)\left(\!\frac{v_{\rm A}}{c}\!\right)
    \left(\!\frac{\delta B}{B_0}\!\right)^{\!\!3}\!\left(\!\frac{l_{\rm esc}}{l _0}\!\right), & 
    \label{eq:Aion}\\
    \ell & \sim 4\tau_{\rm T}(1 - q_{\rm i})\sigma_{\rm e}\left(\delta B/B_0\right)^3\left(v_{\rm A}/c\right)\left(l_{\rm esc}/l_0\right),
    \label{eq:lion}\\
    t_{\rm IC} / t_{0} & \sim (\overline E_{\rm e}/m_{\rm e}c^2) \left[(1 - q_{\rm i})\sigma_{\rm e} (\delta B/B_0)^{2}\right]^{-1}, &
    \label{eq:coolion}
\end{align}
where the Alfv\' en speed $v_{\rm A} = c [\sigma_{\rm i} / (1 + \sigma_{\rm i})]^{1/2}$ is now 
determined by the ion magnetization $\sigma_{\rm i}$. For subrelativistic
ions $\sigma_{\rm i} \simeq B_0^2 / 4\pi n_{\rm i0}m_{\rm i}c^2 \simeq 
\sigma_{\rm e}(n_{\rm e0} / n_{\rm i0})(m_{\rm e}/m_{\rm i}) \simeq \sigma_{\rm e}m_{\rm e}/m_{\rm i}$. 

A model based on an electron-ion composition 
presents a set of open questions with respect to observations. 
Unless almost all energy goes into ion heating ($1 - q_{\rm i}\ll 1$), we 
require $\sigma_{\rm i} \sim \sigma_{\rm e}m_{\rm e}/m_{\rm i} \ll 1$ for 
typical values of the compactness $\ell \sim$ a few 10. A high radiative compactness implies also 
fast cooling ($t_{\rm IC} < t_0$). It is then expected that the turbulence is subject to radiative damping and a significant 
fraction of its power is lost before the energy cascades to the plasma microscales, which is reminiscent of the bulk Comptonization scenario.
However, it is not obvious how the electrons can be maintained at energies of 
about 100 keV if their bulk motions with $\delta v \sim v_{\rm A}$ are nonrelativistic (since $\sigma_{\rm i}\ll 1$) 
and only a small amount of the cascade power arrives at kinetic scales, where the heating usually occurs. 
The issue can be, in principle, avoided if one postulates a rapid form of electron energization that bypasses the turbulent cascade and 
draws energy from all scales of the turbulent flow. 
Another possibility is to assume that the large-scale 
turbulent motions are not constrained by $v_{\rm A}$ (i.e., that the motions are super-Alfv\' enic). 
Finally, mildly relativistic bulk motions and reasonable values of the compactness
can be obtained if $\sigma_{\rm i}\sim \sigma_{\rm e}m_{\rm e}/m_{\rm i} \sim 1$ and  $10^{-3} \lesssim 1 - q_{\rm i} \lesssim 10^{-2}$.

In conclusion,  the nature of turbulent Comptonization in electron-ion plasmas could differ from that 
in pair plasmas. An important parameter affecting the Comptonization is the fraction of turbulence power channeled into ion heating, 
which needs to be investigated with dedicated radiative PIC simulations.

\section{Dependence on the system size}

It is worth commenting on how the limited scale separation in 
our fiducial simulation ($L = 2l_{\rm esc} = 640 d_{\rm e0}$) might affect the results. The observed 
emission is essentially determined by the electron energy distribution and by the optical depth, which
sets the average number of scatterings  
experienced by a photon before escape. The effective electron 
temperature (Eq.~(2) in the main Letter), which controls the position of the Comptonized peak, has no explicit dependence on system size,
although we cannot rule out a moderate implicit dependence. This leaves in question the shape of the electron distribution, in particular its 
nonthermal tail, which controls the gamma-ray emission.

Particles injected into the nonthermal tail in high-$\sigma_{\rm e}$ kinetic turbulence are rapidly accelerated by the nonideal 
electric fields up to $\gamma \sim$ a few $\sigma_{\rm e}$  \cite{Comisso2018}, and not much beyond that if subsequent acceleration is slow compared 
to the cooling time scale \cite{Nattila2021,Sobacchi2021}. Fast acceleration is, however, still possible provided that particles experience relatively 
coherent large-scale fields over their acceleration history. 
Such fields can be, for instance, present in large-scale turbulent 
reconnection layers (e.g., \cite{DalPino2005,Kowal2012,Lazarian2012,Zhang2021,Zhang2023,Chernoglazov2023}), 
where particles on both sides of the reconnecting sheet sample an ideal upstream electric field.
If electrons in black-hole coronae experience fast acceleration across a broad range of scales, their maximum Lorentz 
factor is higher than predicted in our fiducial PIC simulation, and the coronal gamma-ray emission extends to higher energies.

\section{Choice of seed photon distribution}

The choice of the typical seed photon energy, $\overline E_0$, is for a given source constrained by observations. It can be roughly 
identified with the low-energy range of the Comptonized spectrum, which is near 1 keV for Cyg X-1. What remains to be specified is
the shape of the seed photon distribution. Here, we employ a Planck spectrum with temperature $T_0/m_{\rm e}c^2 = 10^{-3}$. This is consistent 
with the common view that the seed photons originate from optically thick and colder regions of the accretion flow, where the
plasma and radiation are near thermal equilibrium \cite{Zdziarski2004}, although seed photons may be additionally provided by 
synchrotron emission \cite{Poutanen2014}.
The energy budget of a turbulent cascade in a plasma of moderate 
optical depth is weakly affected by the choice of $T_0$, 
as long as $\overline E_0 \ll \overline E_{\rm e}$. This is because the escaping photons are 
upscattered to energies $\overline E_{\rm esc}\gg \overline E_0$, and so essentially only the energy of the escaping photons 
enters the energy balance between the turbulent driving and radiative cooling.

\section{Choice of the photon to pair density ratio}

The mean density ratio of photons to electrons and positrons, $n_{\rm ph0} / n_{\rm e0}$, is 
an important parameter of our present model. Reasonable values for $n_{\rm ph0} / n_{\rm e0}$
can be inferred from Eq.~(1) of the main Letter, which relates the amplification factor $A$ to the main
parameters of our model. In particular, the terms in Eq.~(1) show that $n_{\rm ph0} / n_{\rm e0}$ 
scales in proportion to $\sigma_{\rm e}v_{\rm A}/c$ for a fixed value of $A$. For a typical
hard state in X-ray binaries with $A\sim 10$ we require $n_{\rm ph0} / n_{\rm e0} \gtrsim 100$
when the corona is strongly magnetized ($\sigma_{\rm e} \gtrsim 1$). The effect of the density ratio
on the energy spectra can be also seen in Fig.~\ref{figsupp:conv}, which presents a set of numerical
convergence checks performed at two different values of $n_{\rm ph0} / n_{\rm e0}$.
Physically, the photon to electron density ratio in the corona is controlled by the global state of accretion and/or 
by the local balance between pair creation, annihilation, photon emission and absorption.

\section{Role of the turbulence driving}

The structure of coronal turbulence in luminous black-hole accretion flows is presently the subject of ongoing 
investigations. A firm understanding will likely require extreme resolution global 3D MHD simulations, 
of similar type as recently presented for low-luminosity sources \cite{Ripperda2022}, where the nature of 
high-energy emission is different (e.g., \cite{RodriguezRamirez2019}).
In our local model, we drive the turbulence by imposing a large-scale 
time-varying external current \citep{TenBarge2014}. This type of driving excites predominantly, though not exclusively, 
Alfv\' enic perturbations (for discussions of different types of 
turbulence driving see Refs.~\citep{Kowal2012b, Kawazura2020, Zhdankin2021b}).
For strong turbulence an appropriate choice of the amplitude is then such that $\delta B\sim B_0$. 
Strong magnetic perturbations are implicitly assumed throughout this work, although the assumption can be relaxed if
necessary. Perturbations with $\delta B\sim B_0$ could originate, for instance, from the large-scale 
twisting and bending of coronal field lines, anchored in the turbulent accretion disk and in the black-hole magnetosphere.
In contrast, low-amplitude magnetic fluctuations ($\delta B \ll B_0$) 
would result in a more ordered and smooth magnetic field line configuration in the corona. Such perturbations 
could be driven by small-scale ($l_0 \ll l_{\rm esc}$) waves and instabilities in the accreting flow. 
The low-amplitude regime of wave turbulence faces a similar challenge as the low-$\sigma_{\rm e}$ 
scenario (see Sec.~\ref{sec:low_sigma}). Namely, the expected electron distributions
are nearly thermal \cite{Nattila2022} and cannot account for the MeV tail of the observed emission \cite{Poutanen2014}.
A model based on low-amplitude turbulent driving of the corona would have to consistently explain where the missing 
MeV tail of the emission comes from and how it is produced, if not in the corona.  

\section{Comptonization via turbulence and/or reconnection}

It is worth commenting on how the scenario put forward in the 
present work differs from earlier PIC studies of Comptonization 
via bulk motions driven by 2D magnetic reconnection \cite{Sironi2020, Sridhar2021, Sridhar2023}. Reconnection in 2D current
sheets proceeds through the formation of non-turbulent plasmoid chains; photons are then Comptonized by the
bulk plasmoid motions. In contrast, in the present 3D simulations bulk Comptonization is mediated by 
turbulent motions spanning a broad range of scales. It should be noted that turbulence and reconnection
in real 3D systems are intrinsically connected (e.g., \citep{Eyink2013, Kadowaki2018, Lazarian2020}). 
Moreover, turbulence in magnetized plasmas is known 
to form current sheets (e.g., \citep{Politano1995, Lazarian1999, Dmitruk2004, Mininni2006, Kowal2009, SantosLima2010, Eyink2011, Eyink2013, Zhdankin2013}), 
which are also observed in our present simulations (Fig.~3 of the main Letter). To understand the details of how the 
energy transfer occurs through the reconnecting layers in 3D turbulent flows is still an 
open question and requires further investigation.

\section{Additional numerical details}
\label{sec:numerics}

The Compton scattering between macroparticles 
in a given collision cell is calculated using a Monte Carlo approach \cite{Haugbolle2013,DelGaudio2020}. 
The method is based on the selection of a random sample of electron-photon (or positron-photon) 
couples that constitute a list of ``candidates'' for the Compton scattering in a given collision cell and at a given
time step. The computational particles from the randomly generated list are then scattered with a given probability, which is proportional to
the scattering cross section and inversely proportional to the size of the random sample.
In black-hole coronae and other radiatively compact
sources \cite{Fabian2015}, the mean number density of physical photons to electron-positron pairs $n_{\rm ph0}/n_{\rm e0}\gg 1$, 
implying that the frequency of binary collisions experienced by an electron (or positron) is greatly enhanced 
compared to a photon, and relatively large samples of electron-photon (or positron-photon) couples are needed 
for the scattering to be properly captured. To this end, we employ a random sampling 
where each electron (or positron) is paired with at least $i\geq 1$ 
different photons, with $i$ large enough to obtain a good statistical sample for 
a given collision cell. In particular, we determine $i$ at every time step and for each cell 
based on the condition that the maximum probability for an electron to scatter with 
a given photon from the sample is less than 10\%. Typical values of $i$ in our fiducial simulation lie between a few and ten.
Compared to the standard Monte Carlo sampling, where each particle is paired at most once per step, 
our method represents essentially a form of adaptive substepping of the elementary time step. 
It is also worth mentioning that, under the physical conditions explored in this work, 
an electron typically experiences an order-unity deflection only after several binary collisions because most scatterings occur in the Thomson regime.

We use the same average number of computational particles for photons as for the electron-positron pairs, 
and therefore the probability of an electron macroparticle to scatter 
is $p_{\rm e} = p_{\rm ph}n_{\rm ph0}/n_{\rm e0}$, where $p_{\rm ph}$ is the scattering probability for the photon. 
We account for the unequal collision probabilities ($p_{\rm e} > p_{\rm ph}$) using a rejection method \cite{Sentoku2008}.
For each couple from the sample, we draw a uniform random number $r\in[0,1)$, determine 
$p_{\rm e}$ and $p_{\rm ph}$, and calculate the new momenta of the electron (or positron) and photon after
scattering if $p_{\rm e} > r$. 
The momentum of the electron is updated to the new value if  $p_{\rm e} > r$, whereas 
the momentum of the photon is updated if $p_{\rm ph} > r$. The rejection method \cite{Sentoku2008} does not conserve energy and momentum per each Monte Carlo collision, but still does so in a statistical sense.
An alternative that conserves energy and momentum per collision involves splitting of particles \cite{Haugbolle2013,DelGaudio2020}, 
followed by occasional particle merging \cite{Vranic2015}. 
The latter is impractical in the regime explored here, since it would require 
very frequent merging throughout the whole simulation.

\begin{figure*}[htb!]
\centering
\includegraphics[width=0.82\textwidth]{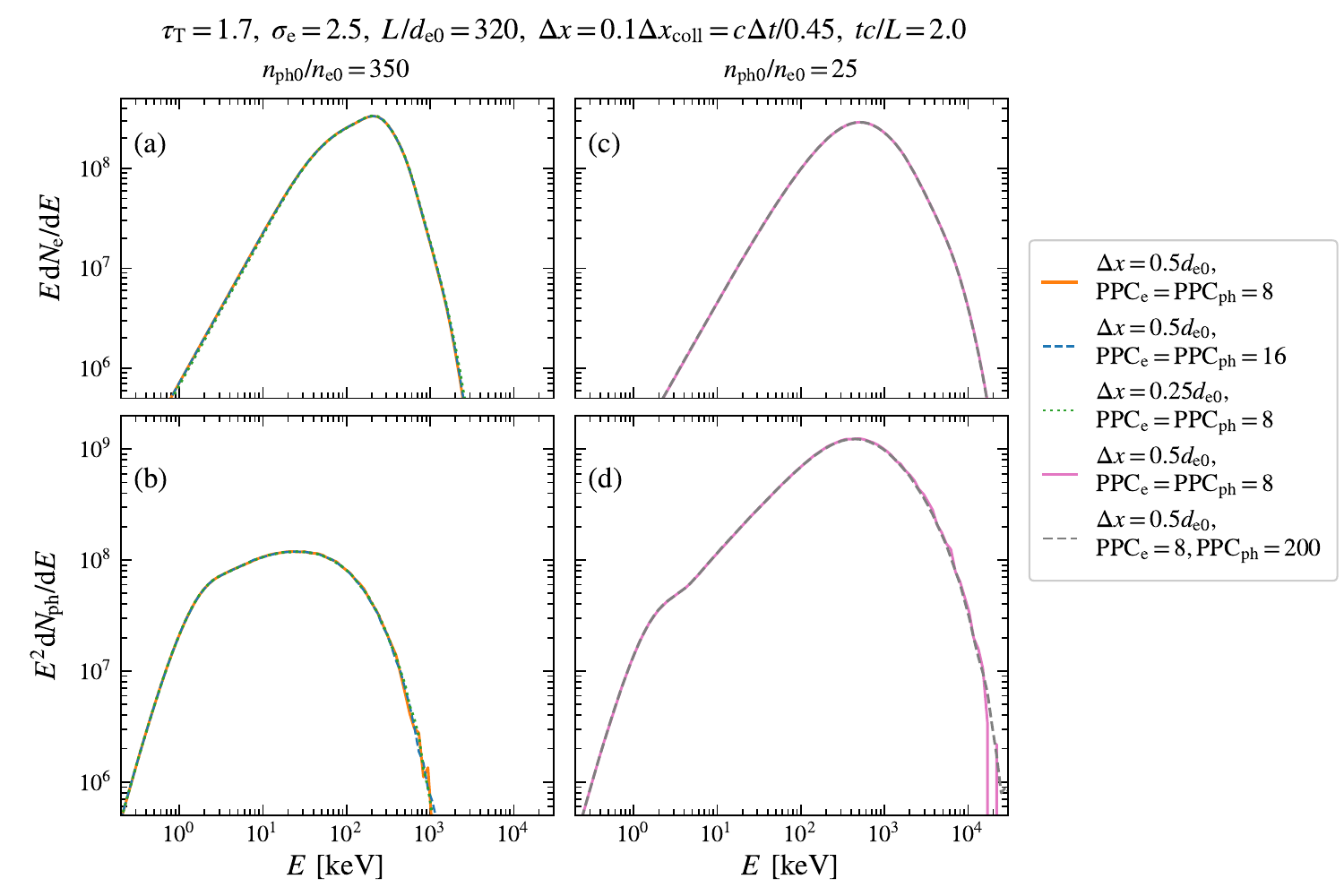}
\caption{\label{figsupp:conv} Dependence of the electron-positron (a,c) and photon (b,d) energy spectra on various numerical parameters (see the main supplement text for details). 
All spectra are shown at around $t = 2 L/c$.}
\end{figure*}

In Fig.~\ref{figsupp:conv} we present a set of numerical convergence checks using a computational box of size $L/d_{\rm e0} = 320$, which is half the size
used in the main Letter. Same as in the main Letter, we set $\tau_{\rm T}=1.7$, $\sigma_{\rm e}=2.5$, and 
$\Delta x = 0.1 \Delta x_{\rm coll} = c\Delta t/0.45$, where $\Delta x$ is the size of a PIC grid cell, 
$\Delta x_{\rm coll}$ is the size of a collision cell for scattering, and $\Delta t$ is the time step.
In panels \hyperref[figsupp:conv]{\ref*{figsupp:conv}(a)-\ref*{figsupp:conv}(b)} we use $n_{\rm ph0}/n_{\rm e0}=350$ and 
check the dependence on the number of particles per cell of the PIC grid ($\rm PPC_{\rm e}$ for pairs and $\rm PPC_{\rm ph}$ for photons), 
and on the spatial resolution.  Our reference simulation with $\Delta x = 0.5 d_{\rm e0}$ and $\rm PPC_{\rm e} = \rm PPC_{\rm ph} = 8$ 
(same as used in the main Letter) is then compared against a simulation with $\rm PPC_{\rm e} = \rm PPC_{\rm ph} = 16$ and another one
where $\Delta x$, $\Delta x_{\rm coll}$, and $\Delta t$ are all twice smaller ($\Delta x = 0.25 d_{\rm e0}$).
In panels \hyperref[figsupp:conv]{\ref*{figsupp:conv}(c)-\ref*{figsupp:conv}(d)} we use $n_{\rm ph0}/n_{\rm e0}=25$ and compare the results obtained for
$\rm PPC_{\rm e} = \rm PPC_{\rm ph} = 8$ against a simulation where  $\rm PPC_{\rm e} = 8$ but $\rm PPC_{\rm ph} = 25\cdot 8 = 200$, such that 
${\rm PPC}_{\rm ph} / {\rm PPC}_{\rm e} = n_{\rm ph0}/n_{\rm e0}$. We use here a more moderate value for $n_{\rm ph0}/n_{\rm e0}$ due to 
memory limitations imposed by the choice ${\rm PPC}_{\rm ph} / {\rm PPC}_{\rm e} = n_{\rm ph0}/n_{\rm e0}\gg 1$. 
When the latter condition is satisfied, the computational electrons and photons are scattered
with equal probabilities ($p_{\rm e} = p_{\rm ph}$) because each macro-photon represents the same number of physical particles
as a macro-electron. The energy and momentum in the simulation with ${\rm PPC}_{\rm ph} / {\rm PPC}_{\rm e} = n_{\rm ph0}/n_{\rm e0}$
are thus conserved in each Monte Carlo collision, which can be compared against the results obtained using the rejection 
method with $\rm PPC_{\rm e} = \rm PPC_{\rm ph}$, where the conservation only holds in a statistical sense.
We find that all electron-positron and photon spectra shown in panels \hyperref[figsupp:conv]{\ref*{figsupp:conv}(a)-\ref*{figsupp:conv}(b)} and 
\hyperref[figsupp:conv]{\ref*{figsupp:conv}(c)-\ref*{figsupp:conv}(d)} are in excellent agreement and conclude that for our typical choice of numerical parameters the
results are well converged.

\end{document}